\documentclass[twocolumn,showpacs,preprintnumbers,amsmath,amssymb]{revtex4}
\usepackage{graphicx}% Include figure files
\usepackage{dcolumn}% Align table columns on decimal point
\usepackage{bm}% bold math
\usepackage{epsfig}

\begin{document}
\title{Accurate extraction of the News}
\author{ Shrirang S. Deshingkar}
\email[]{shrir@hri.res.in}
\affiliation{ 
Harish-Chandra Research Institute
         Chhatnag Road, Jhunsi, Allahabad 211 019, India
}
\begin{abstract}
We propose a new scheme for extracting gravitational radiation from a
characteristic numerical simulation of a spacetime. This method is 
similar in conception to our earlier work but analytical and numerical
implementation is different. The scheme is based on direct transformation 
to the Bondi coordinates and the gravitational waves are extracted by 
calculating the Bondi news function in Bondi coordinates. The entire
calculation is done in a way which will make the implementation easy when 
we use uniform Bondi angular grid at $\mathcal I^+$. Using uniform Bondi grid
for news calculation has added advantage that we have to solve only
ordinary differential equations instead of partial differential equation.
For the test problems this new scheme allows us to extract 
gravitational radiation much more accurately than the previous schemes. 
\end{abstract}


\maketitle

\section{Introduction}
\label{s-intro}

One of the main aims of the numerical relativity simulations of spacetimes
is to calculate the gravitational radiation emitted by various kinds of
sources, particularly in the strong field regime.  The predicted waveforms
for different kinds of astrophysical sources will be useful for detection
of gravitational waves and parameter estimation of sources (by matched
filtering technique) using many working and upcoming gravitational wave
detectors like, LIGO, VIRGO, TAMA, GEO and LISA. For this we need the
simulations to be very accurate and we also need to extract the
gravitational waves faithfully and accurately from the simulations.  Both
these tasks are very challenging.  Binary black holes and neutron stars
are the main likely sources for early detection. In the recent times there
is considerable progress in evolving the binary black hole
\cite{orb_brug1,orb_seidel1,pretorious1, camp1,centrella1, centrella2,
pretorius2,laguna1,centrella3} and binary neutron star \cite{shibata2006}
spacetimes, but a lot of improvements are needed for the predicted
waveforms to be useful for the future observations.

	In the most common approach to the numerical relativity
(Cauchy/ADM approach) the spacetime is foliated by a sequence of spacelike
hypersurfaces parameterized by time. In such a case we cannot extract the
gravitational waves exactly and some approximate techniques have to be
developed.  These techniques  \cite{regw,zeri,monc,rupar,cams,seis} 
use the data at the outer boundary of the simulations to predict the 
gravitational wave content of spacetimes at future null infinity 
($\mathcal I^+$).  In recent times new promising ways,
which are also useful for extracting gravitational waves from Cauchy
simulations of spacetimes have been proposed \cite{bbprl, kins1, kins2, bbs3,
kins4, kinscamp, kins5}. But in this paper we use the characteristic (null
cone)  formulation using the Bondi-Sachs coordinates. The spacetime is
foliated by series of null hypersurfaces parameterized by outgoing
(/ingoing) null geodesics.  Using compactification techniques, $\mathcal
I^+$ is included in the computational domain.  In characteristic numerical
simulations one can in principle extract the gravitational radiation
exactly at null infinity (${\mathcal I}^+$) using the Bondi news function
\cite{bondi,gerochnews,aanews}.  Even then the accurate extraction of the
news seems quite tricky as many numerical complications make the task very
hard.  The news function takes a very simple form in the Bondi
coordinates, but the numerical simulations use more general Bondi-Sachs
coordinates.  The Bondi coordinates correspond to inertial observers at
$\mathcal I^+$, but in a numerical code we want to provide boundary
conditions at inner boundary (e.g. black hole horizon) and then solve the
equations as we go out to $\mathcal I^+$, so we use the more general
Bondi-Sachs coordinates. Earlier news was computed directly in Bondi-Sachs
coordinates \cite{hpn} and then a transformation was done to the Bondi
coordinates (as detectors are more or less in the reference frame of
inertial observer at ${\mathcal I}^+$).  The news expression in
Bondi-Sachs coordinates is quite complicated.

	We had proposed a scheme \cite{news2} where we calculated the news
by going directly to the Bondi coordinates. But further tests showed that
improvement in the news calculations were needed. In certain cases
numerically intermediate constraint, $\tilde J_0 =0 $ (it basically says,
the tangential part of Bondi metric at $\mathcal I^+$ should be zero, the
notation is introduced in Sec. \ref{s-not}) was not getting satisfied,
though as such the final news calculations gave satisfactory result.

	The problem was traced to a particular (i.e. $\eth X$) term in
equation (20)  for $\tilde J_0$ in \cite{news2}. This term is basically
related to an angular derivative of the Bondi angular coordinate with
respect to the Bondi-Sachs angular coordinate. Many techniques, like doing
entire numerical calculations using spin weighted spherical harmonic
coefficients etc. were tried to improve the accuracy of the result while
sticking to the original analytical procedure in \cite{news2}, but it was
hard to achieve the desired results.

	The main reason for this problem seems to be the interaction of
various numerical errors creating unexpected enhancement of error while
calculating angular derivatives. Using the scheme in \cite{news2}, these
angular derivatives had to be calculated in a complicated way, using
Jacobian of coordinate transformation.  This was needed as all the angular
derivatives were with respect to the Bondi-Sachs coordinates while for
calculating the news we needed to use uniform Bondi angular grid. This
choice had to made so that one could interpolate things more easily and
also as then one needed to solve only ordinary differential equations for
doing the coordinate transformation instead of partial differential
equations.

	In this paper we present a scheme which suites these numerical
requirements better and gives the desired accuracy. The scheme is similar
in conception to the scheme in \cite{news2}, but the analytical and
numerical implementation is different. The news is calculated by directly
going to the Bondi coordinates. At the initial slice $J_0$ has to be zero
(notation is introduced in section Sec. \ref{s-not}) and the Bondi-Sachs
and Bondi coordinates have to match at $\mathcal I^+$.  For the news
calculation we stick to the Bondi coordinates and uniform Bondi grid at
${\mathcal I}^+$.  The transformation is done analytically (partly using
Mathematica) so that the numerical implementation will have good accuracy.
We have implemented our scheme and tested it with analytical solutions.
The calculations and results are presented in this paper.  This new scheme
shows very good accuracy for the news extraction and also the constraints
$\tilde J_0 = 0$ is satisfied with desired accuracy.

	The paper starts with a summary of relevant results and notation
for the characteristic formulation of numerical relativity
(Sec.~\ref{s-not}). The coordinate transformation and algebraic
calculations and expressions are given in Sec.~\ref{s-coo}, and then the
procedure for computing the news, at both analytic and computational
levels, is described in Sec.~\ref{s-proc}. The computational tests and
results are presented in Sec.~\ref{s-com} followed by discussions and
conclusions in Sec.~\ref{s-con}.

\section{Notation}
\label{s-not}

	Here we briefly review the notation and formalism for the
characteristic numerical relativity ~\cite{hpn,cce} (see
also~\cite{ntb93,ntb90,rai83,bondi}). The formalism is based on a family
of outing null hypersurfaces using the Bondi-Sachs coordinates. The
hypersurfaces are labeled by $u$, $x^A$ $(A=2,3)$ label the null rays and
$r$ is a surface area coordinate. The metric in these Bondi-Sachs
coordinates~\cite{bondi,sachs} is written as, 
\begin{eqnarray}
   ds^2  =  -\left(e^{2\beta}(1 + {W \over r}) -r^2h_{AB}U^AU^B\right)du^2
\nonumber \\
        - 2e^{2\beta}dudr -2r^2 h_{AB}U^Bdudx^A 
        +  r^2h_{AB}dx^Adx^B,
\label{eq:bmet}
\end{eqnarray}
where $h^{AB}h_{BC}=\delta^A_C$ and
$det(h_{AB})=det(q_{AB})$, with $q_{AB}$ a unit sphere metric.
We use stereographic coordinates $x^A=(q,p)$ for which the unit sphere
metric is
\begin{equation}
q_{AB} dx^A dx^B = \frac{4}{P^2}(dq^2+dp^2),
\end{equation}
where
\begin{equation}
        P=1+q^2+p^2.
\end{equation}
We introduce a complex dyad $q_A$ defined by
\begin{equation}
      q^A=\frac{P}{2}(1,i), \;\;q_A=\frac{2}{P}(1,i)
\end{equation}
with $i=\sqrt{-1}$. For an arbitrary Bondi-Sachs metric,
$h_{AB}$ can then be represented by its dyad component
\begin{equation}
J=h_{AB}q^Aq^B/2.
\end{equation}
$h_{AB}$ 
is uniquely determined by $J$, since the
determinant condition ($det(h_{AB})=det(q_{AB})$) 
implies that there are only two independent 
components of $h_{AB}$ and the remaining dyad component
\begin{equation}
K=h_{AB}q^A \bar q^B /2
\end{equation}
satisfies $1=K^2-J\bar J$. For spherically symmetric case  $J$ is 
identically zero.  We introduce the spin-weighted field,
\begin{equation}
U=U^Aq_A.
\end{equation}
We also introduce complex angular coordinate $\zeta = q+i p $ 
as well as the (complex differential) eth operators $\eth$ and $\bar \eth$
(see~\cite{eth} for full details) which are given as,
\begin{equation}
\eth A = P\frac{\partial A}{\partial {\bar \zeta}} + s A \zeta, \; \;
{\bar \eth} A = P\frac{\partial A}{\partial \zeta} - s A {\bar \zeta},
\label{eq:eth}
\end{equation}
where $A$ is any spin weighted field with spin weight $s$.

The news is calculated  in a conformally compactified coordinates.
Specifically, $(u,r,x^A) \rightarrow (u,\ell,x^A)$ where $\ell = 1/r$. In
$(u,\ell,x^A)$ coordinates, the compactified metric is $d\hat{s}^2=\ell^2 ds^2$
where $\ell$ is a conformal factor with future null infinity ${\mathcal I}^+$
given by $\ell=0$. The compactified metric will be denoted by
$\hat{g}^{\alpha\beta}$ ($\alpha, \beta = 0-3$) and the general compactified 
Bondi-Sachs metric is
\begin{eqnarray}
\hat{g}^{11}=e^{-2\beta}V_a , \;\; \hat{g}^{1A}=e^{-2\beta} U^A,
\nonumber \\
\hat{g}^{10}=e^{-2\beta}, \;\;
\hat{g}^{AB}=h^{AB}, \;\; \hat{g}^{0A}=\hat{g}^{00}=0,
\label{e-bsc}
\end{eqnarray}
where $V_a=\ell^2(1+\ell W)$. In addition to the general compactified
Bondi-Sachs metric Eq.~(\ref{e-bsc}), we will refer to the compactified
Bondi-Sachs metric satisfying the Bondi conditions, and such quantities will
be denoted with a tilde ($\tilde{\;} $), with compactified metric 
and coordinates $\tilde{g}^{\alpha\beta}$ and 
$(\tilde u,\tilde \ell,\tilde x^A)$,
respectively. On ${\mathcal I}^+$, i.e. $\tilde \ell=0$,
$\tilde {g}^{\alpha\beta}$ satisfies the Bondi conditions
\begin{equation}
\tilde {g}_{00}=0, \;\;\tilde {g}_{0A}=0, \;\;\tilde{g}_{01}=1, \;\;
\tilde {g}_{AB}=\tilde q_{AB},
\label{e-b}
\end{equation}
where $\tilde q^{AB}$ is a unit sphere metric with respect to the Bondi angular
coordinates $\tilde x^A$.

\section{Coordinate transformation}
\label{s-coo}

	We define a coordinate transformation, near ${\mathcal I^+}$,
between $(u,\ell,x^A)$ and $(\tilde u,\tilde \ell,\tilde x^A)$ as a Taylor
series expansion in $\tilde \ell$.  Up to second order in $\tilde \ell$ we
can write,
\begin{eqnarray}
u=u_{0} +  A^u \tilde \ell +C^u {\tilde \ell}^2,\;\; 
\ell = \tilde \ell/ \omega + C^\ell {\tilde \ell}^2,
\nonumber
\\ x^A= x^A_{0} +A^A  \tilde \ell  +C^A {\tilde \ell}^2, 
\end{eqnarray}
where $\omega$, $x^A_{0}$, $A^A$, $u_{0}$, $A^u$ are all functions of
${\tilde x}^A$ and $\tilde u$ only. It will turn out that the 
$C^\alpha$ are not needed for the news calculation, but they are
included so that the Jacobian for coordinate transformation 
is manifestly correct to first order in $\tilde \ell$. We also 
introduce complex quantities
\begin{equation}
A={\tilde q}_A A^A \mbox{ and } X = {\tilde q}_A {\tilde x}^A_{0}.
\end{equation}
The metric functions are expanded to first order in $\tilde \ell$ as,
\begin{eqnarray}
\beta= \beta_0 +\tilde \ell \beta_{\tilde \ell}, \;\; 
U=U_0+ \tilde \ell U_{\tilde \ell}, \;\;
J=J_0+ \tilde \ell J_{\tilde \ell}, \nonumber \\
 K=K_0+ \tilde \ell K_{\tilde \ell}, \;\; {\tilde V}_a= \tilde \ell 
V_{a \tilde \ell}.
\end{eqnarray}
\label{eq-ay}
The same first-order expansion in $\tilde \ell$ and notation will be used
for the metric quantities in the coordinate system satisfying the Bondi
conditions (e.g., $\tilde J= \tilde J_0+ \tilde \ell \tilde J_{\tilde
\ell}$). The general compactified Bondi-Sachs metric, and the Bondi-Sachs
metric satisfying the Bondi conditions, are related by
\begin{equation}
({\tilde g_{\alpha\beta})_0}+ \tilde \ell ({\tilde{g}_{\alpha\beta}}
)_{\tilde \ell}=
{\tilde{g}_{\alpha\beta}} = \omega^{2} 
\frac{\partial x^\mu} {\partial \tilde x^\alpha} 
\frac{\partial x^\nu} {\partial \tilde x^\beta} \hat{g}_{\mu\nu},
\end{equation}
with the factor $\omega^{2}$ appearing because there is also an implicit change
of compactification factor from $\ell$ to $\tilde \ell$.

We use Mathematica to find $\tilde {g}^{\alpha\beta}$ to first order in 
$\tilde \ell $, and in doing so we have imposed the conditions,
\begin{equation}
\label{eq:uord}
\frac{\partial{u}}{\partial{\tilde{u}}} = e^{-2\beta_0}
\end{equation}
and
\begin{equation}
\label{eq:xAord}
\frac{\partial{x^A}}{\partial{\tilde{u}}} = U^A e^{-2\beta}/\omega_0.
\end{equation}
These conditions are  equivalent of satisfying,
\begin{equation}
(\partial_u + U^B_0 \partial_B) \tilde x_{0}^A=0
\label{eq-yevol}
\end{equation}
and
\begin{equation}
(\partial_u + U^B_0 \partial_B) \tilde u_{0} = \omega e^{2\beta_0},
\label{eq-uevol}
\end{equation}
along the generators of $\mathcal I^+$ (see  Eqs. (39) and (40)
in~\cite{hpn} for some details). This ensures Bondi conditions
$\tilde g_{00} =0$, $\tilde g_{0A} = 0$, $\tilde g_{01} = 1$ are satisfied
identically. 

	We also evolve $\omega$ along the generators of $\mathcal I^+$ by
integrating \cite{hpn},
\begin{equation}
\label{eq:omegaord}
\frac{d\omega}{du} = -\frac{1}{4} \omega (\eth U + \bar \eth U).
\end{equation}
Evolving $\omega$ using this ordinary differential equation reduces
the errors due to angular derivatives as compared to the method used
in \cite{news2}.  Knowing the coordinate transformation and $\omega$, 
we can find the value of $\tilde J_0$ ($J_0$ in Bondi coordinates),
\begin{eqnarray}
\label{eq:J_B0}
\tilde J_0 & = & -\frac{\omega^2}{4 P^2} ((\eth {\bar X} + B_X {\tilde P} {\bar X} )^2 J_0 
\tilde P^2 + (\eth X)^2 \bar J_0 
    \tilde P^2  \nonumber \\
 &&         +  P^2 (\eth  u)^2 (J_0 {\bar U_0}^2 +   2 K_0 \bar U_0 U_0 
+  {\bar J_0} {U_0}^2)                        \nonumber \\ 
     &&     -2 P \tilde P  \eth  u (J_0  \bar U_0 + K_0 U_0) 
	( {\eth {\bar X}} + B_X {\tilde P} \bar X)  \nonumber \\
   &&       - 2 P \tilde P \eth  u \eth X 
	(K_0 \bar U_0 + \bar J_0 U_0) \nonumber \\
   && + 2 {\tilde P}^2 K_0 {\eth X} ({\eth {\bar X}} + B_X {\tilde P} 
	{\bar X}) ),
\end{eqnarray}
Where, $B_X = \frac{ 2 \tilde \zeta}{\tilde P}$.
The analytical value of $\tilde J_0$ has to be zero as we are in Bondi 
coordinates and we can use the numerical value to estimate the accuracy 
of the coordinate transformation.

	Bondi-Sachs conditions 

\begin{equation}
{\hat{g}^{00} =0},
\end{equation}
gives us from the lowest ($0^{th}$) order in $\tilde \ell $,
\begin{equation}
\label{eq:Au}
A_u =  -\frac{1}{2} \omega e^{2\beta} \eth { u} \bar \eth { u}. 
\end{equation}
First order in $\tilde \ell$ will give us $C_u$ but, it is not needed
for news calculation.
Bondi-Sachs condition,
\begin{equation}
{\hat{g}^{0A} =0}  ,
\end{equation}
leads to,
\begin{equation}
\label{eq:A}
A = -\frac{ \omega e^{2 \beta}}{2 {\tilde P}}
( ( (\bar \eth X + {\bar B_X} {\tilde P} X) {\tilde P}
- {{\bar \eth} u} U_0 P) \eth  u  + {{\bar \eth} u}  \eth X \tilde P  ) 
\end{equation}
from the lowest ($0^{th}$) order in $\tilde \ell $, and will give $C$
from $1^{st}$ order, but it is not needed for our calculation. One can also
get $A_u$ and $A$ by solving for ${\tilde{g}_{11} =0}$ and 
${\tilde{g}_{0A} =0}$
but we use ${\hat{g}^{00} =0}$, ${\hat{g}^{0A} =0}$ as it simplifies the
calculation.

After knowing these quantities, we can calculate $\tilde J_{\tilde \ell}$.
The expression for $\tilde J_{\tilde \ell}$ is quite long and is given in
the equation (\ref{eq:J_Bl}) in the appendix.

\section{Calculating the news}
\label{s-proc}

	After knowing the coordinate transformations to required order we
can develop a systematic scheme for extracting the news. First we describe
the analytic aspects and then we discuss some details about the actual
implementation in the program. It is assumed that the Bondi-Sachs metric
is knows throughout the spacetime.

\subsection{Procedure}
\begin{enumerate}
\item We integrate equation (\ref{eq:xAord}) to get $x^A$ on a fixed Bondi
grid.
\item  We evolve equation (\ref{eq:omegaord}) for $\omega$.
\item $\tilde J_0$  is evaluated form equation  (\ref{eq:J_B0}) and is used 
to monitor the numerical accuracy of the calculation.
\item We find the time transformation using equation (\ref{eq:uord}).
\item $A_u$ is calculated using equation (\ref{eq:Au}) and
$A$ from equation (\ref{eq:A}).
\item We evaluate $\tilde J_{\tilde \ell}$ using equation (\ref{eq:J_Bl}). 
\item The news is extracted by differentiating $\tilde J_{\tilde \ell}$,
\begin{equation}
  N = \frac{1}{2} \frac{\partial^2 \tilde J}{\partial {\tilde \ell} 
	\partial{\tilde u}}
    = \frac{1}{2} \frac{\partial \tilde J_{\tilde \ell}}
       {\partial{\tilde u}}
\end{equation}

\end{enumerate}

\subsection{Computational implementation}

	The news module has been written to interface directly with the
null gravity code in its current form~\cite{roberto}. Thus we use a
compactified radial coordinate $x=r/(R+r)$, with $R=1$. There are $n_x$
points in the $x$ direction in the range [0.5,~1] (corresponding to
$1<r<\infty $). We use stereographic angular coordinates with two patches
with minimum patch overlap of 5 grid points. Both patches have $n_n$ grid
points in each angular direction.

	In actual numerical implementation for the news calculation,
we use uniform ($\tilde u, \tilde \ell, \tilde x^A$) grid instead of 
uniform ($u, \tilde \ell, \tilde x^A$) to avoid having interpolation
with respect to time. This means in place of equation (\ref{eq:uord}) we  
solve its inverse,
\begin{equation}
\frac{d \tilde u}{du} = \omega e^{2\beta},
\end{equation}
along the generators of $\mathcal I^+$.  This also means that in
the derivative operators with Bondi coordinates, have to be replaced by, 
\begin{equation}
\frac {\partial} {\partial {\tilde u}} \rightarrow 
\frac{e^{-2\beta_0}}{\omega} \frac {\partial} {\partial u}
\end{equation}
and
\begin{equation}
\frac {\partial} {\partial {\tilde x}^A}|_{\tilde u} \rightarrow
\frac {\partial} {\partial {\tilde x}^A}|_{u}
+ \frac{e^{-2\beta_0}}{\omega}  \frac {\partial {\tilde u}} {\partial 
{\tilde x}^A}|_{u}
\frac {\partial} {\partial u}.
\end{equation}

	We also introduce $\bar \eth U$ and $\eth U$ as well as $\bar \eth
X$ and $\eth X$ as additional variables. This helps to reduce the
numerical errors as typically news indirectly involves $3^{rd}$ or higher
angular derivative of $U$ and $X$ (if the original coordinates are not
already Bondi coordinates). Higher angular derivatives are known to give
lot of problems numerically. Introducing these additional variables is one
simple way of improving the numerical behavior. These variables should be
introduced in the null evolution code as well. All integration and
differentiation schemes are $2^{nd}$ order, apart from for the first step,
for which time integration is done by $1^{st}$ order scheme.

\section{Computational tests and results}
\label{s-com}

We implemented and tested our method for calculating the Bondi news
function for Schwarzschild and linearized Robinson-Trautman in rotating as
well as tumbling coordinates.  The analytical solution on the discretized
grid is given as the input to the news module. The news module as such
sees it as input from an equivalent numerical evolution code. The same
news module can be used with suitable null evolution code.

\subsection{Schwarzschild solution in rotating and tumbling coordinates}

We get the Schwarzschild solution in Bondi-Sachs rotating coordinates
by transforming the standard angular coordinates by, 
$\phi \rightarrow \tilde \phi + \kappa u$. Various metric coefficients at
$\mathcal I^+$ are written in this case as,
\begin{center}
\begin{eqnarray}
W_N = W_S = 0, &
\beta_N = \beta_S = 0, \nonumber \\
J_N = J_S = 0, & 
(J_{\ell})_N = (J_{\ell})_S = 0, \nonumber \\
U_N = -\frac{ 2 i \kappa \zeta_N}{M(1 + \zeta_N \bar \zeta_N)}, & 
U_S = \frac{ 2 i \kappa \zeta_S}{M(1 + \zeta_S \bar \zeta_S)},
\end{eqnarray}
\end{center}
where the subscripts N and S refer to north and south patch.

	When we have coordinate system which is rotating around an
equatorial axis, we call it a tumbling coordinate system. In this case the
relationship of the tumbling angular coordinates to the Bondi angular
coordinates is given as,
\begin{equation}
\tilde \theta = \arccos(\cos{\theta} \cos{\kappa u} +
                        \sin{\theta} \sin{\phi}\sin{\kappa u})
\end{equation}
\begin{equation}
\tilde \phi = \arctan {\left ( \frac{\cos {\kappa u} \sin \theta \sin \phi -
             \cos\theta \sin{\kappa u}} {\sin \theta \cos \phi}\right )}
\end{equation}
In these Bondi-Sachs coordinates only nonzero part of the metric at 
$\mathcal I^+$ is $U$, which is given as,
\begin{eqnarray}
U_N = i \kappa \frac{1- {\zeta_N}^2 }{(1 + \zeta_N \bar \zeta_N)}, \nonumber \\
U_S = i \kappa \frac{1- {\zeta_S}^2 }{(1 + \zeta_S \bar \zeta_S)} .
\end{eqnarray}

The convergence tests were performed for the following grid sizes for all
the tests,
\begin{eqnarray}
(a)& & \Delta q=\Delta p = 1/8, \Delta x = 1/64,  \Delta u = 0.04 \nonumber \\
(b)& & \Delta q=\Delta p = 1/16, \Delta x = 1/128, \Delta u = 0.02 \nonumber \\
(c)& & \Delta q=\Delta p = 1/32, \Delta x = 1/256, \Delta u = 0.01 \nonumber \\
(d)& & \Delta q=\Delta p = 1/64, \Delta x = 1/64, \Delta u = 0.005.
\label{e-grids}
\end{eqnarray}
For the highest resolution run we keep $\Delta x = 1/64$ due to memory 
limitations, but as such the results do not change with $\Delta x$.

\begin{figure}[!]
\epsfig{file=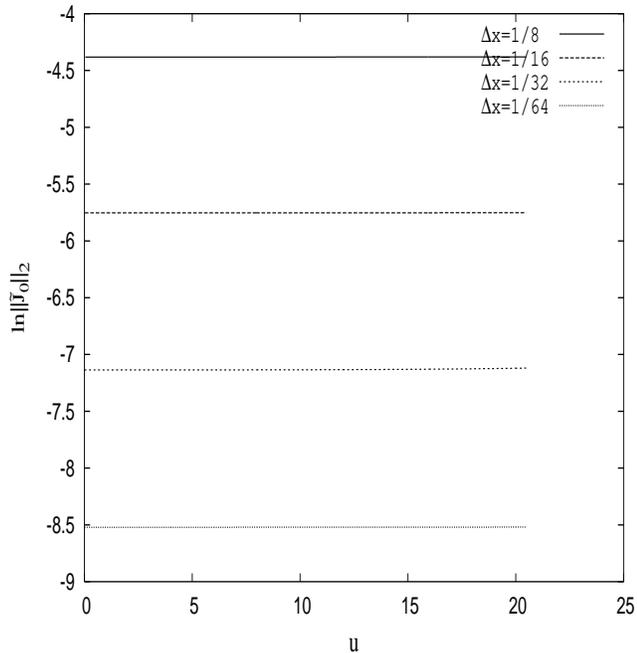,height=3.4in,width=3.5in,angle=-90}
\caption{$L_2$ norm of $||\tilde J_{0}||_2$  for 
Schwarzschild in tumbling coordinates as function of $u$. The norm
changes very slowly with $u$. Analytically $\tilde J_{0}$ should be zero. 
The graph for Schwarzschild in rotating coordinates looks essentially 
identical, so we have not plotted it separately.}
\label{fig:TuJR0}
\end{figure}

In both these cases the  $L_2$ norm of $\tilde J_0$ shows 2nd (1.99) order 
convergence to zero as in figure
(\ref{fig:TuJR0}) 
and the convergence rate is essentially independent of $u$. 

	The news function with our scheme remains identically zero. This
is an ideal behavior and is a pleasant surprise. The news shows this
behavior as $A$ and $A_u$ are zero analytically and with our scheme and
choice of variables they also remain zero numerically. From the expression
for $\tilde J_\ell$ one can see that it will be zero for our case. Due to
this we can expect to extract out news accurately even if we happen to be
in a rotating or tumbling kind of background coordinates and the amplitude
of gravitational radiation is very small.

	The scheme in \cite{news2} showed required convergence for $\tilde
J_0$ in rotating Schwarzschild case, but later it was realized it had
problems while dealing with the tumbling Schwarzschild case. It was very
hard to remove these numerical problems in that scheme. Also the error in
the news in both these cases, though, convergent to $2^{nd}$ order, they
were not all that small, if one had to extract very weak gravitational
wave signals in nontrivial coordinates. Both these issues lead us to
develop the scheme presented above.

\subsection{Linearized Robinson-Trautman solution}
\label{s-rt}

	The Robinson-Trautman solution represents a distorted black hole
emitting purely outgoing radiation. The radiation decays exponentially,
and asymptotically the solution becomes Schwarzschild. In the linearized
case when the amplitude of the perturbation (i.e. also the gravitational
radiation) is small one can write it as addition of different spherical
harmonics ($Y_{lm}$) of various amplitudes (see e.g. \cite{news2}).

\subsubsection{Rotating coordinates}

	The metric components for the linearized Robertson Trautmann
solution, when only $Y_{20}$ term is present (i.e. only the $l=2$, $m=0$
term of spherical harmonics is present in the perturbation) can be written
at $\mathcal I^+$ as \cite{yosefthesis},
\begin{center}
\begin{eqnarray}
W_N  = W_S & = & 0,  \nonumber \\
\beta_N  = \beta_S & = & 0, \nonumber \\
J_N  = J_S & = & 0,  \nonumber \\
(J_{\ell})_N & = & -12\lambda_{20}e^{-2u/M} M \frac{ {\zeta^2}_N}
{(1 + \zeta_N \bar \zeta_N)^2}, \nonumber \\
(J_{\ell})_S & = &-12\lambda_{20}e^{-2u/M} M \frac{ {\zeta^2}_S}
{(1 + \zeta_S \bar \zeta_S)^2}, \nonumber \\
U_N & = & -\frac{ 2 i \kappa \zeta_N}{M(1 + \zeta_N \bar \zeta_N)}, \nonumber \\
U_S & = & \frac{ 2 i \kappa \zeta_S}{M(1 + \zeta_S \bar \zeta_S)},
\end{eqnarray}
\end{center}
where the subscripts $N$ and $S$ refer to north and south patch, $M$ gives
the mass of the remnant black hole and $\lambda_{20}$ is the initial (at 
$u=0$) amplitude of the perturbation.

In this case $\tilde J_0$ shows $2^{nd}$ order convergence to zero.
Basically the $\tilde J_0$ part remains same as for the corresponding 
Schwarzschild case (Fig. \ref{fig:TuJR0}). The news also shows $2^{nd}$ 
order convergence to the analytical value on the Bondi grid given by,
\begin{equation}
N = -12 \lambda_{20} e^{-2u} \frac{\tilde \zeta^2}{(1 + 
\tilde \zeta \bar {\tilde \zeta})^2},
\label{eq:RTNews}
\end{equation}
on both the patches.

\begin{figure}[!]
\epsfig{file=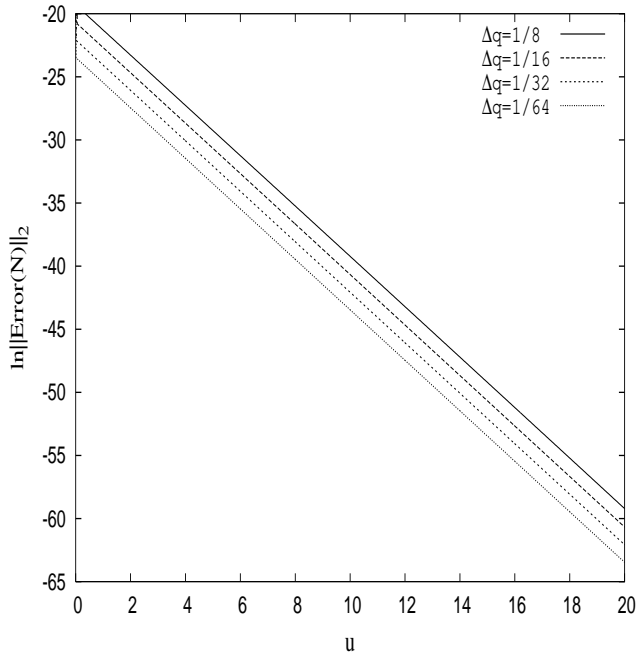,height=3.4in,width=3.5in,angle=-90}
\caption{ $L_2$ norm of the error in the news as a function of time $u$ for
various discretizations for the linearized Robinson-Trautman solution in
rotating coordinates. }
\label{fig:NewsRRT}
\end{figure}

For our test in figure (\ref{fig:NewsRRT}) we set $ M =1$, 
$\lambda_{20} = 3\times {10}^{-7}$ and $\kappa =0.01$.
Even for very small or very hight value of $\lambda_{20}$, 
news shows  $2^{nd}$ order
convergence though the gravitational wave amplitude goes down by more than
$10$ order of magnitude as the time passes. The convergence rate are
shown in table (\ref{tab:ConvTuRT}).
This is much better behavior 
than we could get by the earlier news modules. 

\begin{table}[t]
\begin{tabular} {|c|c|c|c|}
\hline

convergence rate at $u $ & grids (a) \& (b) & (b) \& (c) & (c) \& (d) \\
\hline

4 & 2.012 & 2.002 & 2.008 \\
\hline

8 & 2.028 & 2.021 & 2.005 \\

\hline

12 & 2.070 & 2.013 & 2.005 \\
\hline

16 & 2.097 & 2.012 & 2.005 \\

\hline

20 & 2.086 & 2.011 & 2.005 \\

\hline

\end{tabular}

\caption{Change of convergence rate with $u$ for the $L_2$ norm of error
in the news function for linearized Robinson-Trautman in rotating coordinates 
for different grid resolutions in equation \ref{e-grids}.}

\label{tab:ConvRoRT}
\end{table}

\subsubsection{Tumbling coordinates}

\begin{figure}[!]
\epsfig{file=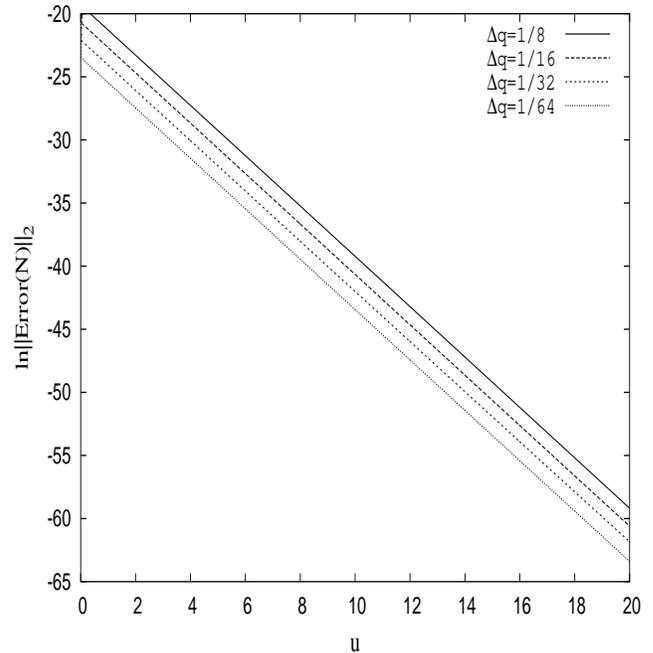,height=3.4in,width=3.5in,angle=-90}
\caption{ $L_2$ norm of the error in the news as a function of time $u$ for
various discretizations for the linearized Robinson-Trautman solution in
tumbling coordinates. }
\label{fig:NewsTRT}
\end{figure}

	Various metric components for the linearized Robertson Trautmann
solution, when only $Y_{20}$ perturbation term is present can be written
at $\mathcal I^+$ in tumbling coordinates \cite{yosefthesis} as,
\begin{center}
\begin{eqnarray}
W_N &=& W_S = 0, \nonumber \\
\beta_N &=& \beta_S = 0, \nonumber \\
J_N & = & J_S = 0, \nonumber \\
(J_{\ell})_N & = & 3 M \lambda_{20}e^{-2u/M}  \frac{ 2 i \zeta_N \cos {\kappa u}
+ ( 1+{\zeta_N}^2) \sin\kappa u)^2}
{(1 + \zeta_N \bar \zeta_N)^2}, \nonumber \\
(J_{\ell})_S & = & 3 M \lambda_{20}e^{-2u/M}  \frac{ 2 i \zeta_S \cos {\kappa u}
+ ( 1+{\zeta_S}^2) \sin\kappa u)^2}
{(1 + \zeta_S \bar \zeta_S)^2}, \nonumber \\
U_N & = & i \kappa \frac{1- {\zeta_N}^2 }{(1 + \zeta_N \bar \zeta_N)}, \nonumber \\
U_S & = & i \kappa \frac{1- {\zeta_S}^2 }{(1 + \zeta_S \bar \zeta_S)} .
\end{eqnarray}
\end{center}
In this case we again set $M=1$, $\lambda_{20} = 3 \times 10^{-7}$ and
$\kappa =0.01$. With this data, $\tilde J_0$ shows $2^{nd}$ order
convergence to zero as for the Schwarzschild in tumbling coordinates.
$L_2$ norm of the error in the news also shows $2^{nd}$ order convergence
to zero as shown in figure (\ref{fig:NewsTRT}) and table
(\ref{tab:ConvTuRT}).  The accuracy of news extraction remains very good
as long as equator of one patch in Bondi coordinates doesn't go over the
pole of the other patch in Bondi-Sachs coordinates due to tumbling. As
such the accuracy of the news extraction does not depend on the initial
amplitude of the waveform even it is increased or decreased by more that
10 orders of magnitude.

\begin{table}[t]
\begin{tabular} {|c|c|c|c|}
\hline

convergence rate at $u $ & grids (a) \& (b) & (b) \& (c) & (c) \& (d) \\
\hline

4 & 2.015 & 2.002 & 2.002 \\
\hline

8 & 2.036 & 1.982 & 2.032 \\

\hline

12 & 2.055 & 1.943 & 2.064 \\
\hline

16 & 2.049 & 1.906 & 2.106 \\

\hline

20 & 2.031 & 1.862 & 2.156 \\

\hline

\end{tabular}
\caption{Change of convergence rate with $u$ for the $L_2$ norm of  
error in the news function for linearized Robinson-Trautman in 
tumbling coordinates for different grid resolutions in equation \ref{e-grids}.}

\label{tab:ConvTuRT}
\end{table}

\section{Conclusions}
\label{s-con}

	We have developed a new scheme for extracting the news from
characteristic numerical simulations of the spacetimes. Like our earlier
scheme this scheme is based on a coordinate transformation from general
Bondi-Sachs coordinates to Bondi coordinates. But in the present scheme
the transformation is done in a different way. It is done so that we can
easily use a uniform and constant Bondi angular grid for the news
calculation. This has an advantage that while transforming coordinates one
has to evolve ordinary differential equations instead of partial ones.
Also the interpolation and angular differentiation becomes easier and more
accurate. We also introduce $\eth U$ and $\bar \eth U$ as auxiliary
variables in the code to overcome problems related to numerical
differentiations.

	Like earlier schemes the present scheme also needs $J=0$ at
$\mathcal I^+$ on the initial time ($u=0$) slice. It also needs the
Bondi-Sachs and Bondi coordinates to match on the initial slice. We can
also check the accuracy of coordinate transformation by monitoring values
of $\tilde J_0$, which should be zero analytically.

	Our new scheme shows $2^{nd}$ order convergence of ${\tilde J_0}$
to zero and overcomes the convergence problem of ${\tilde J_0}$ faced by
our earlier scheme for the Schwarzschild solution in tumbling coordinates.  
Also with this scheme the news for Schwarzschild solution in
rotating/tumbling coordinates stays exactly zero, while with the earlier
scheme it is not so small value which was converging to zero to $2^{nd}$
order in grid size. In other cases also our new scheme shows $2^{nd}$
order convergence in extracting the news. For the Robinson-Trautman test,
it successfully extracts out the news even when the amplitude of the
gravitational waves is very small and we are in a coordinate system which
is rotating/tumbling with respect to the Bondi coordinates. The news
calculation remains accurate even though the signal goes down many orders
of magnitude with time.  The relative accuracy of the news extraction does
not depend on the initial amplitude of the waveform even when it is
increased or decreased by more that 10 orders of magnitude from the cases
studied in this paper. This indicates our scheme gives very accurate and
faithful results and the scheme is quite robust.

	Introducing $\eth U$ and $\bar \eth U$ as auxiliary variables
seems to play a crucial role in getting the news accurately.  In most
cases, whichever scheme we may choose to calculate the news, it will
indirectly contain at least third angular derivatives of $U$. Higher
angular derivatives are known to create problems in the null code. To take
care of this problem earlier three complex variables corresponding to
$\eth \beta$, $ \bar \eth J$ and $ \eth K$ were introduced in
\cite{gomez}. In a similar manner we suggest to introduce eth derivatives
of $U$ as new variables. It will be particularly useful when one is
extracting gravitational waves from the simulations. All the tests we have
done show significant improvement if we introduce these new variables.

	The accuracy of our new scheme is very encouraging and we expect
that the present method will be useful for extracting gravitational
radiation from more realistic physical systems/simulations.  It could play
an important role in Cauchy-Characteristic extraction. We plan to explore
this direction in a future work.

\newpage
\begin{widetext}
\appendix
\section{Equation for $\tilde J_{\ell}$} \label{ap:ap1}
The expression $J_{\ell}$ can be written as,
\begin{eqnarray}
{\tilde J_{\ell}} & = & (-4 \eth u K_0
          \omega( (\eth X(\bar J_0 - K_0) {\tilde P} +
(-J_0 + K_0) \tilde P (\eth {\bar X} + {\tilde P }B_X {\bar X}) +
 2i P {\eth u} {\mathcal Im} ({\bar U_0}(J_0 -K_0))   \nonumber \\
&&  +  P{\eth u} K_0 U_0 ) (-2 P (  U_\ell -{\bar U_\ell} + A_u \omega  
(U_u - {\bar U_u})) +   
            2i {\mathcal Im} ({\bar A} \omega {\tilde P} (-\eth U +
 \eth {\bar  U } + {\tilde P} \bar U_0 X)))\nonumber \\
&&  + (-{\tilde P}(J_0 +K_0)({\eth {\bar X}} + {\tilde P} B_X 
{\bar X})
- {\tilde P} {\eth X} ({\bar J_0} + K_0)
  + 2 P {\eth u}{\mathcal Re}({\bar U_0}(J_0 + K_0)  )) \nonumber \\
&& ((2 P ((U_{\ell} +\bar U_{\ell}) +  A_u \omega
   (U_u + \bar U_u) ))
        + 2 {\mathcal Re}( \bar A \omega {\tilde P} (\eth U +
\eth {\bar U} + {\tilde P} {\bar U_0} X)))) \nonumber \\
&& - (8 \omega ( ({\eth {\bar X}} + B_X {\tilde P} {\bar X})^2 
K_0 {\tilde P}^2 +
                  P^2 ({\eth u})^2 {\bar  U_0} (K_0 {\bar U_0}
+ {\bar J_0} U_0) +
                  ({\eth {\bar X}} + B_X {\tilde P} {\bar X}) {\tilde P}^2 
{\eth X} {\bar J_0}  \nonumber \\
&& -  {\tilde P} P {\eth u}    ({\eth{\bar X}} + B_X {\tilde P} {\bar X}) 
( 2 K_0 {\bar U_0} +  {\bar J_0} U_0 ) +
     P {\eth u} {\tilde P} (-\eth X {\bar J_0} {\bar U_0} )) 
 (P (J_{\ell} + A_u {J_0}_{,u} \omega) \nonumber \\
&& + \frac{ \omega {\tilde P}}{4} ({\bar A} ( 2 {\eth J} + (1 -i) J_0 {\tilde P}
                 {\bar X} 
 -  ( 1+i ) J_0 {\tilde P} X)
      +  A (2 {\bar \eth J} +  (1+i)  J_0 {\tilde P} {\bar X} - (1-i)
          J_0 {\tilde P} X))))/ P \nonumber \\
&& - (8 \omega ( (\eth X)^2 K_0 {\tilde P}^2 
+ ({\eth {\bar X}} + B_X {\tilde P} {\bar X}) J_0 {\tilde P} 
({\eth X} {\tilde P} - P {\eth u} U_0) +
                  P {\eth u}
  U_0 (P {\eth u}  (J_0 {\bar U_0} + K_0 U_0) ) \nonumber \\
&& + {\eth X} (P {\eth u} {\tilde P}(-J_0 {\bar U_0} - 2 K_0 U_0) )) 
(P ({\bar J_{\ell}}
        + A_u {\bar J_0},_u \omega) +
\frac{ \omega {\tilde P}}{4}(A (2 {\bar \eth}{\bar J} -  (1+i) {\bar J_0}
                {\tilde P}
                   {\bar X} +  (1-i) {\bar J_0} {\tilde P} X) \nonumber \\
&& +  {\bar A} (2 \eth {\bar J}  -  (1-i)
                        {\bar J_0} {\tilde P}
               {\bar X} +  (1+i) {\bar J_0} {\tilde P} X))))/P +
        16 K_0 ( P^2 \eth u (2 e^{2 \beta} {\eth \omega}
- {\eth A_u} \omega^2 (J_0 {\bar U_0}^2 + U_0 (2 K_0 {\bar U_0} +
                        {\bar J_0} U_0)))  \nonumber \\
&& - \omega^2 {\tilde P}^2 (({\eth {\bar A}} + {\bar A} B_X \tilde P)
( {\eth {\bar X}} J_0 + B_X J_0 {\tilde P} {\bar X}+{\eth X} K_0) 
 + {\eth A} ({\eth X} {\bar J_0} + {\eth {\bar X}} K_0 +
                        B_X K_0 {\tilde P}{\bar X} )) \nonumber \\
&& + P \omega^2   {\tilde P} (({\eth u} ( {\eth {\bar A}} + 
B_X {\tilde P} {\bar A})  (J_0 {\bar U_0} + K_0 U_0) +
        {\eth A} (K_0 {\bar U_0} +  {\bar J_0} U_0)) + 
     {\eth A_u} (({\eth {\bar X}} +  B_X {\tilde P} {\bar X})
(J_0 {\bar U_0} + K_0 U_0) \nonumber \\
&& + {\eth X} (K_0 {\bar U_0} +  {\bar J_0} U_0) )) 
+ P^2 e^{2\beta }{(\eth u)}^2 \omega
                V_{a\ell}))/(16 P^2 K_0)
\label{eq:J_Bl}
\end{eqnarray}
\end{widetext}

\bibliography{news4i}

\end{document}